# LUMINESCENCE PROPERTIES OF A FIBONACCI PHOTONIC QUASICRYSTAL

**Vasilios Passias, Nikesh Valappil, Zhou Shi, Lev Deych,
Alexander Lisyansky and Vinod M. Menon**[*]

*Department of Physics, Queens College of The City University of New York (CUNY)*

## ABSTRACT

We report the realization of an active one-dimensional Fibonacci photonic quasi-crystal via spin coating. Manipulation of the luminescence properties of an organic dye embedded in the quasi-crystal is presented and compared to theoretical simulations. The luminescence occurs via the pseudo-bandedge mode and follows the dispersion properties of the Fibonacci crystal. Time resolved luminescence measurement of the active structure shows faster spontaneous emission rate, indicating the effect of the large photon densities available at the bandedge due to the presence of critically localized states. The experimental results are in excellent agreement with the theoretical calculations.

---
[*] Corresponding Author: vmenon@qc.cuny.edu

Light can be manipulated in desirable ways by allowing it to interact with a medium whose refractive index is modulated over length scales corresponding to the light's wavelength. The refractive index modulation may be periodic, random or quasi periodic. Since the behavior of a photon in a periodic structure is analogous to that of an electron in a semiconductor crystal, such structures are called photonic crystals. Since their discovery, periodic photonic crystals have received much attention due to the numerous fundamental and technological applications [1-8]. In these crystals, Bragg scattering and refraction happens due to constructive interference leading to the formation of photonic bandgaps similar to electronic bandgaps in solids. One of the major breakthroughs in this field has been the realization of ultralow threshold lasers using active photonic crystals consisting of either quantum wells or quantum dots embedded in them [9, 10]. This was possible due to the development of high quality factor cavities with small mode volumes permitting the modification of spontaneous emission.

At the other extreme from these periodic systems are completely random structures which display novel light transport and localization properties. These disordered systems have demonstrated a plethora of phenomena such as Anderson localization, coherent back scattering, optical Hall Effect, universal conductance fluctuations of light waves, random lasers, localization in resonant media and levy statistics among others.

While there have been numerous studies on light propagation and the nature of light emission from completely periodic and random structures, very little is known about the transport of light and emission characteristics of deterministic but non-periodic structures [11]. Such structures, called quasicrystals, may exist in one-, two-, or three-dimensional realizations and exhibit properties intermediate between periodic and completely disordered systems. Among one-dimensional quasicrystals the most studied are structures based on the Fibonacci sequence. Both classical (light) and quantum (electrons) waves in these structures have been shown to have a self-similar energy spectrum [12], a pseudo bandgap corresponding to regions of forbidden frequencies [13], and critically localized states whose wave functions are characterized by power law asymptotes and self-similarity [14]. All these properties are typical for quasi-periodic structures so that Fibonacci superlattices are considered as archetype structures to study the fundamental aspects of wave propagation in quasicrystals [15].

The first experimental study of Fibonacci structures was concerned with electron transport in semiconductor heterostructures [16] and was followed by numerous other experimental and theoretical works dealing with electron propagation in such systems [17-22]. Light propagation in Fibonacci quasicrystals were studied in the seminal work by Gellerman, et al [23], which was followed by studies of dispersion [24], omni-directional reflection [25, 26], and temporal dynamics of pulse propagation in these systems [27].

All the studies performed so far have considerably improved our understanding of light transport in passive Fibonacci quasicrystals. At the same time, structures containing optically active elements arranged in the form of a Fibonacci quasicrystal have not yet received any attention except of a recent theoretical study of reflection spectra of quantum well based Fibonacci lattices [28]. However, effects of critically localized modes existing at the edge of the pseudogap in these structures on their emission properties have yet to be studied either theoretically or experimentally. In this work we report the realization and study of luminescence of an active one dimensional Fibonacci photonic quasi-crystal based on polymer layers containing embedded organic dye molecules.

The one-dimensional photonic crystal was built by stacking two different materials A and B and was designed using the following deterministic generation scheme: $S_{j+1} = \{S_{j-1}, S_j\}$ for j ≥1, where $S_0 = \{B\}$ and $S_1 = \{A\}$. For j≥1 the layer sequences are: $S_2 = \{S_0 S_1\} = \{BA\}$; $S_3 = \{S_1 S_2\} = \{ABA\}$; $S_4 = \{S_2 S_3\} = \{BAABA\}$, etc. In the present work, the two material chosen were {A} - Cellulose Acetate (CA) and {B} - Poly Vinyl Carbazole (PVK). The choice of the two polymers was dictated by their refractive index contrast and their solubility. The refractive indices of CA and PVK at 600 nm are 1.475 and 1.683, respectively. CA is soluble in polar solvents such as diacetone alcohol and PVK is soluble in non-polar solvent such as chlorobenzene.. Similar approach using alternating layers of spin coated polymers has been previously used to realize Bragg mirrors for bandedge lasing, electro optical switching, enhanced spontaneous emission and microcavity lasing [29-32].

A schematic drawing of the active one-dimensional Fibonacci photonic crystal is shown in Fig. 1. Also shown in Fig. 1 is the simulated transmissivity of the structure calculated using transfer matrix method [33]. A pseudo photonic bandgap is observed around 600-700 nm. In the present work we focused on the bandedge states that lie on the high frequency side of this pseudogap. The electromagnetic field distribution in the structure at the bandedge, shown in the inset of Fig. 1b, clearly indicates a localized bandedge state.

The $S_9$ Fibonacci photonic crystal was realized on a glass substrate via spin coating. The polymers PVK with concentration of 5mg in 20 ml of chlorobenzene and CA with concentration of 30 mg in 10 ml of diacetone alcohol were used to fabricate the structure. The {AA} layer was realized using a CA solution having twice the concentration. The CA polymer solutions were then infiltrated with Sulforhodamine dye, with an emission wavelength of 618 nm. The concentration of the dye in the CA solution was 1 mg in 10 ml of diacetone alcohol.

The fabrication methodology involved spin coating process for every layer of polymer added onto the structure, which was subsequently heated on a hot plate for 15 minutes to remove the solvent. Cellulose acetate layers were heated at 120ºC, while PVK layers were heated at 80ºC. The difference in heating temperatures was required to completely evaporate the solvent of the former and eschew the formation of cracks on the structure. The desired thickness of the CA and PVK layers to yield a pseudo-bandgap around 600 nm were 132 nm and 109 nm and were realized by spin coating the layers at 1800 rpm and 2200 rpm, respectively. The {AA} layer was realized by spin coating the CA solution with double the concentration at 3000 rpm. The layer thicknesses were separately calibrated using reflectivity and surface profiling. Experiments revealed that the emission wavelength of the dye in CA was 594 nm, instead of bare dye's emission wavelength (618 nm). This change introduced smaller overlap between the emission spectra and the bandedge state. Despite this detrimental effect, optical characterization of the system revealed interesting results which are discussed below.

The Fibonacci photonic crystal was characterized using reflectivity and photoluminescence (PL) measurements. All optical measurements reported here were carried out at room temperature. Angular resolved reflectivity measurements were carried out using a fiber coupled Tungsten-Halogen lamp as the white light excitation source. Light from the excitation source was focused to spot size of approximately 1mm in diameter and thus averaging out spatial inhomogeneities arising from spin coating. The reflected light was collected by a fiber coupled spectrometer (Ocean Optics HR-2000 with 600 μm core diameter collection fiber).

Continuous wave PL measurements were carried out using the 488 nm line of an argon ion laser for optical excitation and excitation spot diameter of 1mm. An emission peak is observed at the wavelength corresponding to the higher frequency bandedge state. Figure 2 shows the experimental reflectivity at normal incidence along with the theoretical simulation of such a structure. Also shown in the same figure are the experimentally observed steady state PL spectra and the simulated PL spectra.

Theoretical simulation of the PL spectra was based on the transfer matrix approach developed in Ref. [34, 35]. In this method we describe luminescence by solving Maxwell equations with a random source term imitating non-coherent nature of spontaneous emission. For s-polarized waves, for instance, the Maxwell equation for a component of the field perpendicular to the plane containing growth direction of the structure (designated as $z$ direction) and wave vector **k** takes the form

$$\frac{d^2 E(z)}{dz^2} + q^2 E(z) = -4\pi \frac{\omega^2}{c^2} P_{nc},$$
$$q^2 = \frac{\omega^2}{c^2} \varepsilon(\omega, z) - k^2, \tag{1}$$

where $\omega$ is the frequency of the wave, $k$ is in-plane wave number, $c$ is the speed of light in vacuum, and $\varepsilon(\omega, z)$ is the dielectric constant, whose dependence on $z$-coordinate reflects the quasi-periodic distribution of the layers constituting the structure. When $z$ falls inside a passive PVK layer $\varepsilon(\omega,z) = \varepsilon_B = 1.683$, for values of $z$ inside CA layers embedded with dye the value of the dielectric constant is

$$\varepsilon(\omega,z) = \varepsilon_A + \frac{A}{\omega^2 - \omega_{abs}^2 + 2i\gamma\omega}, \qquad (2)$$

where the second term takes into account the contribution of optical transitions induced in dye molecules into the dielectric constant of the layer *B*. Frequency $\omega_{abs}$ corresponds to the maximum of the absorption of dye molecules dissolved in CA polymer layers, while $\gamma$ characterizes the width of the absorption line. Term $P_{nc}$ on the right hand side of Eq. (1) reflects additional polarization due to spontaneous optical transitions between energy levels of the dye molecules. Respective levels are populated due to incoherent relaxation processes following initial excitation of the molecules in PL experiments. Using, for instance, the density matrix formalism, one can show that this polarization can be described by the following equation

$$\left(\omega^2 - \omega_{em}^2 + 2i\gamma\omega\right) P_{nc}(z,\omega) = \Gamma \omega_{em} \Sigma(z,\omega,k), \qquad (3)$$

where $\Gamma$ is proportional the oscillator strength of the transition responsible for the luminescence, and $\Sigma(z,\omega,k)$ is the non-coherent source of the polarization modeled by a random function. Statistical properties of this function must be determined from microscopic calculations, however, for our purposes it is sufficient to describe these properties by a phenomenological correlation function

$$\left\langle \Sigma(\omega_1,k_1,z_1)\Sigma^*(\omega_2,k_2,z_2) \right\rangle = S\delta(\omega_1-\omega_2)\delta(k_1-k_2)\delta(z_1-z_2). \qquad (4)$$

Writing down Eq. (4) we assumed that the random process responsible for spontaneous emission is uncorrelated in time and space, which is a reasonable assumption given the time and spatial scale typical for our experimental situation. One should note that we introduced two different resonance frequencies $\omega_{em}$ and $\omega_{abs}$ to describe emission and absorption properties of the dye molecules, which, of course, cannot be justified by simple density matrix calculations. This is done in order to reflect a large experimentally observed Stokes shift between emission and absorption spectra of the dye molecules. The origin of this shift lies in an interaction between molecules and phonons. A completely consistent theory of luminescence in this situation would have to include this interaction. However, since our goal is to study effects of the macroscopic photonic structure on the luminescence, it is sufficient to take the Stokes shift into account phenomenologically in the manner reflected in Eqs. (2) and (3).

We solve Eq. (1) with the source term given by Eq. (3) following the approach of Refs. [33, 34], in which the amplitude of the non-coherently emitted light is expressed in terms of transfer-matrices describing transport of light across the structure:

$$E_-^{(m)} = \frac{\langle -|T_R T(N, m+1)|V^{(m)}\rangle}{\langle -|T_{PC}|-\rangle}, \tag{5}$$

where $E_-^{(m)}$ is the amplitude of the light emitted to the right of the structure, two by two transfer matrices $T_R$, $T(N, m+1)$, and $T_{PC}$ describe propagation of light across the interface between the last layer of the structure and vacuum, between the $m^{th}$ and the last layers, and across the entire quasi-crystal structure, respectively. Two dimensional bra and ket vectors appearing in Eq. (5) are given as $|-\rangle = \begin{pmatrix} 0 \\ 1 \end{pmatrix}$, $|V^{(m)}\rangle = \begin{pmatrix} V_1^{(m)} \\ V_2^{(m)} \end{pmatrix}$, where

$$V_1^{(m)} = -\frac{\Gamma \omega_{em}}{\omega^2 - \omega_{em}^2 + 2i\gamma\omega} \frac{e^{iqz_+^{(m)}}}{2q} \int_{z_-^{(m)}}^{z_+^{(m)}} e^{-iqz} \Sigma(z) dz,$$

$$V_2^{(m)} = i\frac{\Gamma \omega_{em}}{\omega^2 - \omega_{em}^2 + 2i\gamma\omega} \frac{e^{-iqz_+^{(m)}}}{2q} \int_{z_-^{(m)}}^{z_+^{(m)}} e^{iqz} \Sigma(z) dz.$$

(6)

and $z_\pm^{(m)}$ are right and left coordinates of $m^{th}$ active layer respectively. Total emitted intensity is given as $I \propto \sum_m \langle |E_-^{(m)}|^2 \rangle$, where brackets indicate averaging over random function $\Sigma$ with correlator given by Eq.(4). Comparison between theoretical and experimental data was carried out by fitting theoretical reflection curves to the experiment adjusting layer thicknesses, and then using the obtained parameters to calculate the luminescence spectra without any additional fitting. In the light of this procedure we can conclude that we obtained an extremely good agreement between theory and experiment in terms of shape and position of the main luminescent peak in the vicinity of the bandedge. This agreement is not a trivial fact because the position of this peak is determined by interplay between luminescence spectra of dye molecules as measured outside of the photonic structure, and the band structure of the quasi-crystal. At the same time, our calculations failed to reproduce fine details of the experimental spectra such as appearance of the secondary peaks. This discrepancy can be explained by several factors such as inhomogeneous broadening of the dye spectra, which has been ignored in our calculations, or the presence of unintentionally formed defect layers in the structure. We studied the role of the former of these factors by taking into account the inhomogeneous broadening in the effective medium approximation. Our calculations showed that indeed, the inhomogeneous broadening results in appearance of the second peak in the spectra.

We also performed angle dependent reflectivity and PL measurements to obtain the dispersion characteristics. Reflectivity measurements were taken in 5° steps from 0° to 60° angles of incidence. The dispersion of the pseudo-bandedge mode obtained through these measurements is shown in Fig. 3 along with the theoretically predicted dispersion curve. The theoretical curve is obtained using TM simulations for different angles.

The possibility to alter the emission characteristics of an emitter by engineering the environment is one of the key fundamental concepts in the field of photonics. Under weak coupling between the emitter and the bandedge mode, the emission follows the bandedge mode dispersion and the emission spectrum takes on the linewidth of the bandedge state. To verify this we performed angle dependent PL measurements with collection from 0° to 20° in 5° steps and the obtained PL peak positions (red squares) are also shown in Fig. 3. Beyond 20° the emission from the dye lies outside the pseudo-bandgap and hence does not follow the dispersion curve. Also shown in Fig. 3 is the theoretically predicted PL peak position (black squares). Both the experimental PL and the reflectivity agree very well with the theoretically predicted dispersion indicating coupling between the emitter and the bandedge state.

To further substantiate this, we performed time resolved PL measurements using a time correlated single photon counting set up (Horiba Jobin Yvon). The active Fibonacci photonic quasicrystal was pumped using a 469 nm diode laser with 50 ps pulse width. The emission from the sample was detected by a high speed photomultiplier tube via a monochromator. Measurements were carried out with the emission monochromator on and off resonance with respect to the pseudo-bandedge mode. All the measurements were carried out with the same concentration of the dye in CA. The results of the time resolved PL measurements are shown in Fig. 4. A control sample consisting of the dye in CA with the same concentration as that used in the Fibonacci structure was used for comparison. The sulforhodhamine dye in CA showed PL lifetime of 4.57 ns. The dye showed decay time of 4.4 ns when observed off resonance from the bandedge state, while on resonance we observed noticeable reduction in the PL lifetime of 3.9 ns. This reduction in PL lifetime is attributed to the increased photon density of states available at the bandedge due to the presence of critically localized states.

In summary, we have successfully demonstrated the fabrication of an active Fibonacci photonic crystal structure with an organic dye embedded in it and demonstrated the modification of luminescence properties. Theoretical modeling based on the transfer-matrix approach confirmed the origin of the main luminescence peak as a result of coupling between dye transitions and the pseudo-gap state of the photonic structure. The entire structure was fabricated using spin coating. The emission from the dye followed the dispersion of the bandedge state and demonstrated spectral narrowing. Time resolved PL measurements on the structure indicate faster radiative decay rate due to the large photon density of states available at the edge of the pseudo bandgap due to the presence of critically localized states. Through better overlap between the bandedge states and the gain maximum of the dye, mirrorless lasing may be achievable in these structures due to the long photon lifetimes and large photon densities supported by the localized modes.

**ACKNOWLEDGEMENTS**

We would like to acknowledge financial support from Air Force Office of Scientific Research (Grant # F49620-02-1-0305), Army Research Office (Grant # W911 NF-07-1-0397) as well as PCS-CUNY grants.

# Figure Captions

**Figure 1**(a) Schematic drawing of the spin-coated one-dimensional Fibonacci photonic quasicrystal and (b) its calculated transmissivity. Shown in the inset of 1(b) is the calculated electric field squared distribution in the structure at the pseudo-bandedge indicating a localized mode.

**Figure 2** Transmissivity and room temperature photoluminescence (PL) observed from the active Fibonacci photonic quasicrystal. These measurements were done at room temperature and the light was collected normal to the surface of the sample. Also shown in the figure are the theoretically simulated transmission and PL spectra. Excellent agreement is found between the theory and experiment.

**Figure 3** Dispersion curve of the pseudo-bandedge mode obtained from angle resolved reflectivity measurements (white squares) along with the photoluminescence (PL) peak (red squares) wavelengths. The blue line is the theoretically predicted dispersion for reflectivity and the black squares indicated the theoretically predicted PL peak position.

**Figure 4** Time resolved photoluminescence spectra of the rhodamine dye embedded in the Fibonacci photonic quasicrystal on resonance (blue) and off resonance (green) with the pseudo bandedge mode. Also shown is the time resolved PL observed from the bare dye (red). Reduction of lifetime or increasesed spontaneous emission rate is observed for the case of dye when the emission is in resonance with the pseudo-bandedge mode due to the large photon density available.

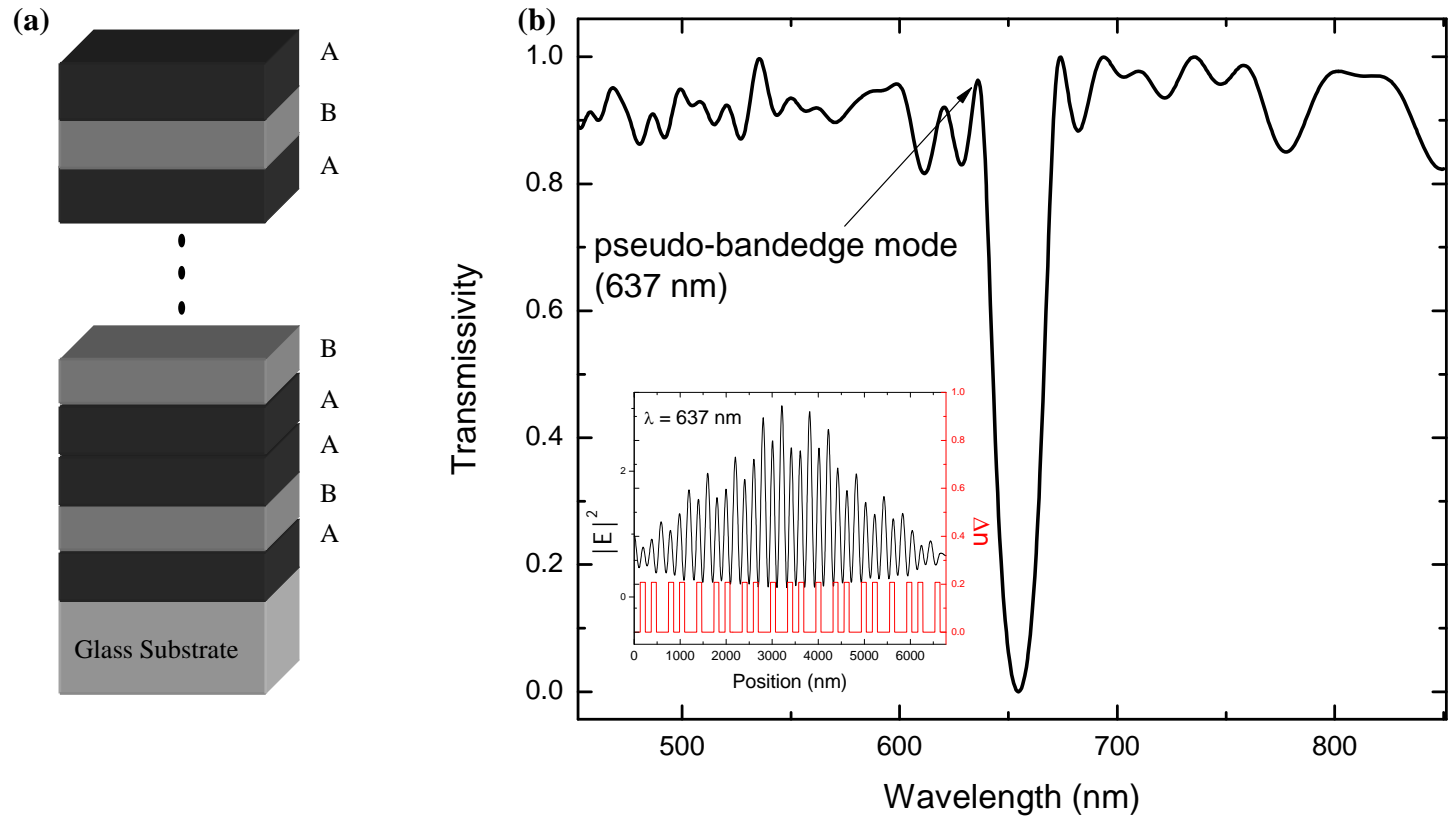

Fig. 1 (a) Schematic drawing of the spin-coated one-dimensional Fibonacci photonic quasicrystal and (b) its calculated transmissivity. Shown in the inset of 1(b) is the calculated electric field squared distribution in the structure at the pseudo-bandedge indicating a localized mode.

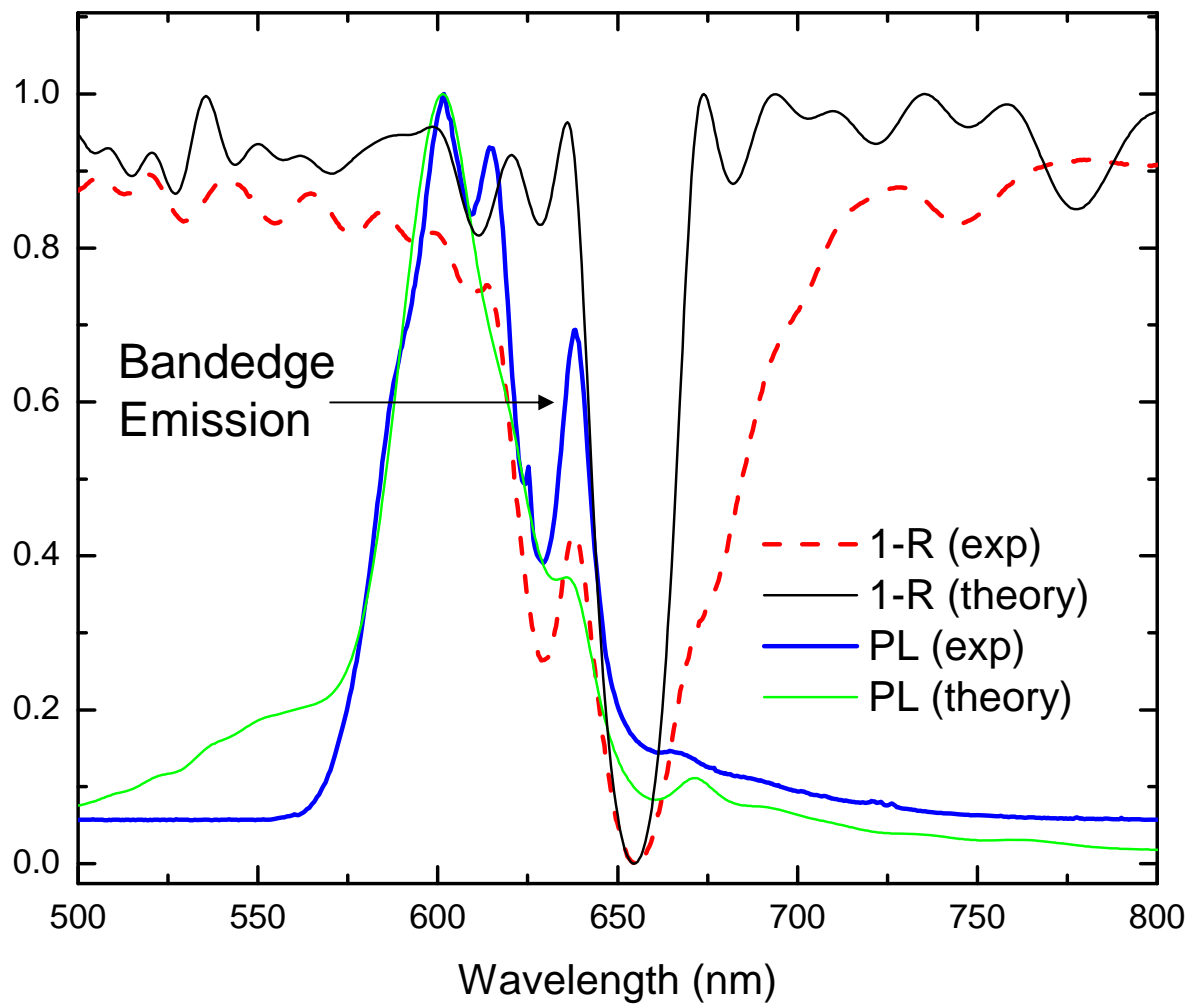

Fig. 2 Transmissivity and room temperature photoluminescence (PL) observed from the active Fibonacci photonic quasicrystal. These measurements were done at room temperature and the light was collected normal to the surface of the sample. Also shown in the figure are the theoretically simulated transmission and PL spectra. Excellent agreement is found between the theory and experiment.

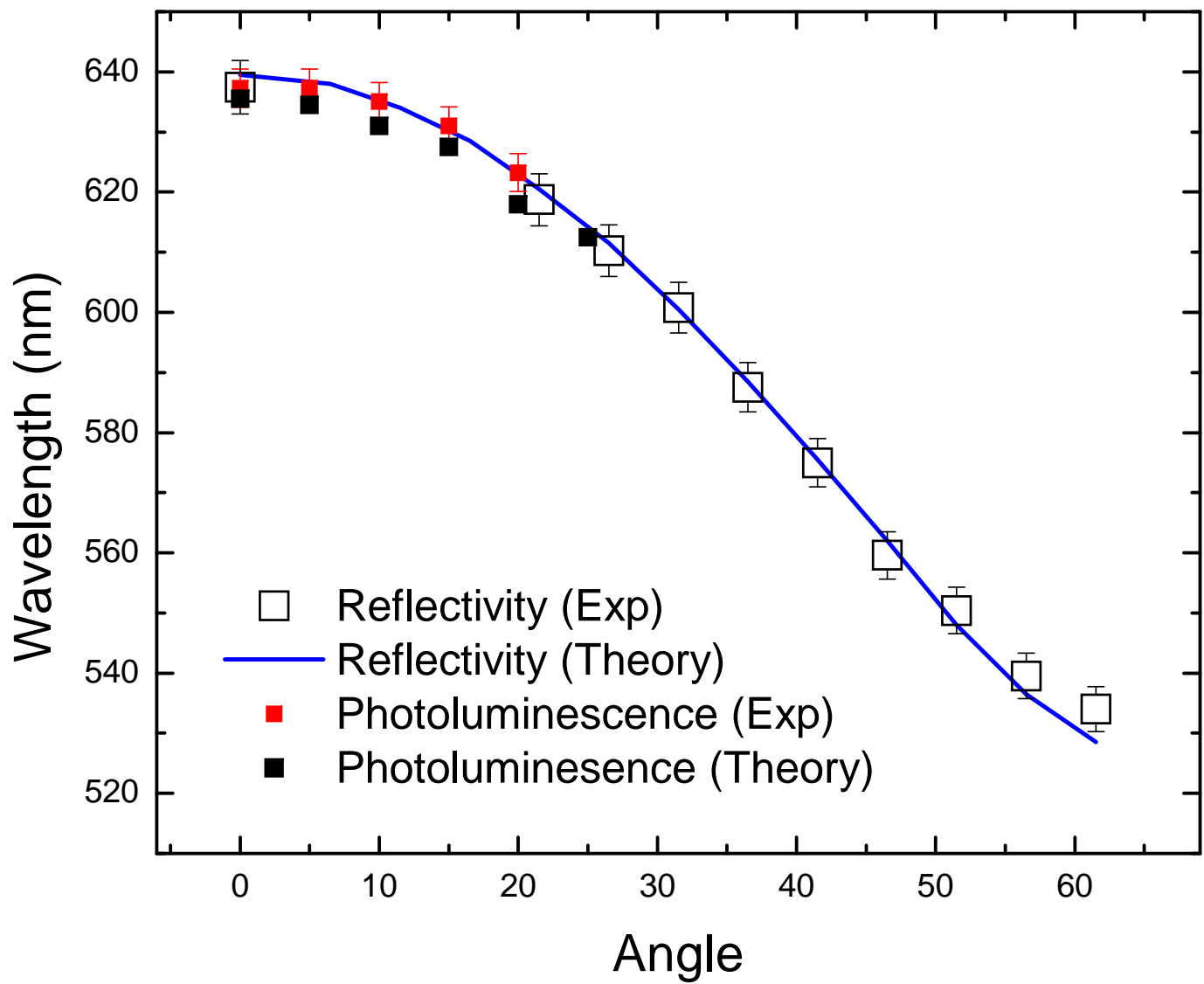

Fig. 3 Dispersion curve of the pseudo-bandedge mode obtained from angle resolved reflectivity measurements (white squares) along with the photoluminescence (PL) peak (red squares) wavelengths. The blue line is the theoretically predicted dispersion for reflectivity and the black squares indicated the theoretically predicted PL peak position.

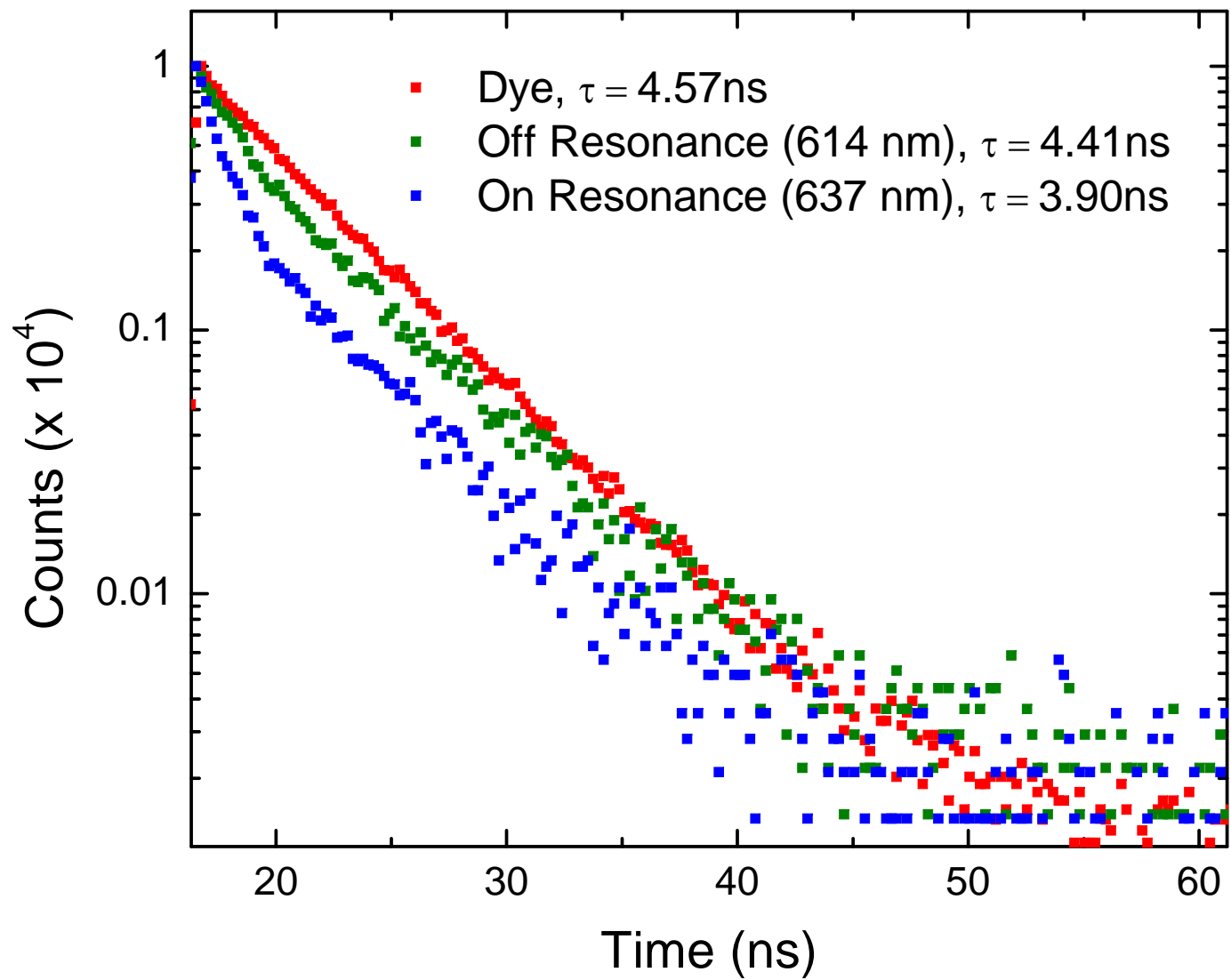

Fig. 4 Time resolved photoluminescence spectra of the rhodhamine dye embedded in the Fibonacci photonic quasicrystal on resonance (blue) and off resonance (green) with the pseudo bandedge mode. Also shown is the time resolved PL observed from the bare dye (red). Reduction of lifetime or increasesed spontaneous emission rate is observed for the case of dye when the emission is in resonance with the pseudo-bandedge mode due to the large photon density available.